\documentclass[fleqn,usenatbib]{mnras}

\usepackage[T1]{fontenc}
\usepackage{lmodern}
\DeclareRobustCommand{\VAN}[3]{#2}
\let\VANthebibliography\thebibliography
\def\thebibliography{\DeclareRobustCommand{\VAN}[3]{##3}\VANthebibliography}

\usepackage{graphicx}	
\usepackage{amsmath,amssymb}	
\usepackage{multirow}
\usepackage{booktabs}

\usepackage{scalerel,tikz}
\usetikzlibrary{svg.path}
\definecolor{orcidlogocol}{HTML}{A6CE39}
\tikzset{orcidlogo/.pic={
 \fill[orcidlogocol] svg{M256,128c0,70.7-57.3,128-128,128C57.3,256,0,198.7,0,128C0,57.3,57.3,0,128,0C198.7,0,256,57.3,256,128z};
 \fill[white] svg{M86.3,186.2H70.9V79.1h15.4v48.4V186.2z}
 svg{M108.9,79.1h41.6c39.6,0,57,28.3,57,53.6c0,27.5-21.5,53.6-56.8,53.6h-41.8V79.1z M124.3,172.4h24.5c34.9,0,42.9-26.5,42.9-39.7c0-21.5-13.7-39.7-43.7-39.7h-23.7V172.4z}
 svg{M88.7,56.8c0,5.5-4.5,10.1-10.1,10.1c-5.6,0-10.1-4.6-10.1-10.1c0-5.6,4.5-10.1,10.1-10.1C84.2,46.7,88.7,51.3,88.7,56.8z};
}}
\newcommand\orcidicon[1]{\href{https://orcid.org/#1}{\mbox{\scalerel*{
\begin{tikzpicture}[yscale=-1,transform shape]
\pic{orcidlogo};
\end{tikzpicture}
}{|}}}}

\newcommand{\kmps}{\mathrm{km\,s^{-1}}}
\newcommand{\tff}{t_\mathrm{ff}}
\newcommand{\otff}{1\,\tff}
\newcommand{\ottff}{1.2\,\tff}
\newcommand{\sigv}{\sigma_v}
\newcommand{\sigvod}{\sigma_{v,\mathrm{1D}}}
\newcommand{\sigvtd}{\sigma_{v,\mathrm{3D}}}
\newcommand{\co}{\mathrm{CO}}
\newcommand{\cooz}{\mathrm{CO~(1-0)}}
\newcommand{\coto}{\mathrm{CO~(2-1)}}
\newcommand{\omegav}{\Omega(v)}
\newcommand{\nth}{^{\mathrm{th}}}
\newcommand{\hyd}{\rm{H_2}}
\newcommand{\mach}{\mathcal{M}}
\newcommand{\rco}{\mathcal{R}_{\mathrm{CO}}}
\newcommand{\rcooz}{\mathcal{R}_{\mathrm{CO,1\!-\!0}}}
\newcommand{\rcoto}{\mathcal{R}_{\mathrm{CO,2\!-\!1}}}
\newcommand{\csf}{\mathcal{C}_{\mathrm{SF}}}

\title[Turbulence from CO observations]{Turbulence inference from CO spectral observations}

\author[Narayan et al.]{
Jayashree Narayan$^{\orcidicon{0009-0007-8280-1010}\,1,2}$,\thanks{E-mail: \href{mailto:jayashreenarayangs@gmail.com}{jayashreenarayangs@gmail.com}}
Aris Tritsis$^{\orcidicon{0000-0003-4987-7754}\,3}$\thanks{E-mail: \href{mailto:aris.tritsis@epfl.ch}{aris.tritsis@epfl.ch}}, and 
Christoph Federrath$^{\orcidicon{0000-0002-0706-2306}\,1}$\thanks{E-mail: \href{mailto:christoph.federrath@anu.edu.au}{christoph.federrath@anu.edu.au}}
\vspace{0.1cm}\\
$^{1}$Research School of Astronomy and Astrophysics, Australian National University, Canberra, ACT~2611, Australia\\
$^{2}$Indian Institute for Science Education and Research, Mohali, Knowledge city, Sector~81, SAS Nagar, Punjab 140306\\
$^{3}$Institute of Physics, Laboratory of Astrophysics, Ecole Polytechnique Federale de Lausanne (EPFL),
    \\\quad Observatoire de Sauverny, CH-1290, Versoix, Switzerland
}

\pubyear{\the\year{}}

\begin{document}
\label{firstpage}
\pagerange{\pageref{firstpage}--\pageref{lastpage}}
\maketitle

\begin{abstract}
Turbulence influences the structure and dynamics of molecular clouds, and plays a key role  in regulating star formation. We therefore need methods to accurately infer turbulence properties of molecular clouds from position-position-velocity (PPV) spectral observations. A previous method calibrated with simulation data exists to recover the 3D turbulent velocity dispersion from PPV data. However, that method relies on ``optically-thin conditions'', ignoring any radiative transfer (RT) and chemical effects. In the present study we determine how opacity, RT, and chemical effects influence turbulence measurements with CO lines. We post-process a chemo-dynamical simulation of a turbulent collapsing cloud with a non-local thermodynamic equilibrium line RT code to generate PPV spectral cubes of the $\cooz$ and $\coto$ lines, and obtain moment maps. We isolate the turbulence in the first-moment maps by using a Gaussian smoothing approach. We compare the CO results with the optically-thin scenario to explore how line excitation and RT impact the turbulence measurements. We find that the turbulent velocity dispersion ($\sigv$) measured via CO requires a correction by a factor $\rco$, with $\rcooz=0.88^{+0.09}_{-0.08}$ for the $\cooz$ line and $\rcoto=0.88^{+0.10}_{-0.08}$ for the $\coto$ line. As a consequence, previous measurements of the $\sigv$ were overestimated by $\sim10-15\%$ on average, with potential overestimates as high as $40\%$, taking the 1-sigma uncertainty into account.
\end{abstract}

\begin{keywords}
ISM: clouds -- ISM: kinematics and dynamics -- ISM: molecules -- radiative transfer -- stars: formation -- turbulence
\end{keywords}



\section{Introduction}
\label{sec:introduction}
Turbulence plays a crucial role in the evolution of molecular clouds and the star formation process, acting as a key mechanism that regulates the dynamics and structure of star-forming regions \citep{2004RvMP...76..125M,McKee_CF,PadoanEtAl2014}. In the absence of turbulence, molecular clouds would rapidly collapse under their own gravity, leading to an unrealistically high star formation rate and efficiency. Turbulence slows down star formation \citep{Federrath2015}, while simultaneously generating localised density enhancements that promote fragmentation and star formation \citep{2004RvMP...76..125M}. The nature of turbulence is thought to set the initial mass function of stars \citep{PadoanNordlund2002,HennebelleChabrier2008,10.1111/j.1365-2966.2012.20730.x,10.1093/mnras/stab505,MathewFederrathSeta2023}, and the star formation rate \citep{KrumholzMcKee2005,PadoanNordlund2011,HennebelleChabrier2011,FederrathKlessen2012}.

Considering the importance of turbulence in the interstellar medium, we require reliable techniques to measure the physical properties of the turbulence. One such key property is the turbulent Mach number, which in turn is governed by the three-dimensional (3D) turbulent velocity dispersion, denoted as $\sigvtd$. An observational key challenge is that we can only measure the line-of-sight (LOS) velocity component through Doppler shifts of spectral lines, while the two transverse velocity components remain inaccessible \citep{2001ApJ...546..980O}, requiring sophisticated methods to estimate and recover the full 3D turbulent velocity distribution of interstellar clouds. \citet{stewart_and_federrath} developed a method to reconstruct the intrinsic 3D turbulent velocity distribution from LOS velocity measurements molecular clouds. Their method was calibrated and tested with simulations of rotating, turbulent, collapsing molecular clouds, providing good estimates of $\sigvtd$ from position-position-velocity (PPV) data.

However, the method by \citet{stewart_and_federrath} assumes that the gas is optically thin. By contrast, PPV observations in molecular clouds rely on molecular lines that are often not optically thin, such as the $^{12}\cooz$ rotational line. Radiative transfer (RT) and excitation effects can have a significant impact on these lines, the inferred moment maps, and ultimately on the estimates of the intrinsic $\sigvtd$. In such cases, radiation is absorbed and scattered before it can reach the observer, potentially masking the cloud’s internal structure \citep{1981MNRAS.194..809L}. This affects the measurement of velocity dispersion and inferred turbulence, as molecular emission lines may become saturated or altered in dense regions, leading to incomplete or skewed data \citep{1976A&A....50..327B}. RT modelling is thus crucial in order to understand systematic biases in measurements of the velocity dispersion, and the overall dynamics of molecular clouds. Here, we propose a method to correct for the effects of finite optical depth, line excitation, and RT, in order to recover the intrinsic $\sigvtd$ from CO line observations.

In Sec.~\ref{sec:methods} we summarise our methods and the numerical tools to perform the chemo-dynamical simulation and the RT calculations. In Sec.~\ref{sec:results} we present our method to isolate turbulence in first-moment maps, and our results on the effect of RT and chemistry in measurements of the turbulent velocity dispersion. Additionally, we perform a power spectrum analysis, and investigate the robustness of our results as the cloud dynamically evolves. Sec.~\ref{sec:effect_on_turb} provides a discussion of the impact of CO excitation and RT on the $\sigvtd$ inference and introduces a correction factor ($\rco$) to account for the effects of CO excitation and RT. Sec.~\ref{sec:conclusions} summarises the main findings of this work.

\section{Methods}
\label{sec:methods}

\subsection{Hydrodynamical simulations}

\label{sec:hydrodyn_sim}
We use the isothermal MHD simulation of a turbulent collapsing molecular cloud described in \cite{TritsisBasuFederrath2025a}. Here, we give a brief overview of the simulation and refer the interested reader to the aforementioned study for more details. The simulation is performed with the adaptive mesh refinement code \textsc{FLASH} \citep{2000ApJS..131..273F, 2008ASPC..385..145D} using open boundary conditions, a base resolution of $64^3$ grid points and two levels of refinement. The physical size of the simulated cloud is $L=2\,\mathrm{pc}$ in each direction, the initial density is $500~\rm{cm^{-3}}$, the initial strength of the magnetic field is $7.5\,\mu \mathrm{G}$, and the temperature is equal to 10 K. Consequently, the cloud is supercritical and its total mass is $\sim240\,\rm{M_\odot}$.

We add turbulent initial conditions using the publicly-available code \texttt{TurbGen} \citep{2010A&A...512A..81F,FederrathEtAl2022ascl}. The velocity field is generated in Fourier space (in the range $2\le k \le 20$ where $k=2\pi/L$), following a power-law dependence of $dv^2/dk \propto k^{-2}$. The velocity field has a standard deviation of $\sigma_v = 0.554\,\mathrm{km/s}$, which, given our sound speed of $c_\mathrm{s}=0.18\,\mathrm{km/s}$, corresponds to a turbulent sonic Mach number of $\mach\approx3$.

Alongside solving the equations of Ideal MHD, we incorporate a non-equilibrium gas-grain chemical network to model the formation and destruction of 115~chemical species. The inputs to our chemical model for every timestep and grid point are the temperature, $\hyd$ number density, 3D visual extinction and the cosmic-ray ionization rate with the source/sink terms calculated after the advection terms. The visual extinction in each grid cell was estimated ``on the fly'' using the six-ray approximation, and the column density of the gas, following \citet{2012MNRAS.421..116G}. To convert between the column density in each direction and the visual extinction, we adopt the relation $N_\mathrm{H_2}/A_\mathrm{v} = 9.4 \times 10^{20}\,\rm{cm^{-2}}\,mag^{-1}$, following \citet{2010ApJ...721..686P}. We assume a constant grain abundance of $n_{g^-}/n_\mathrm{H_2} = 10^{-12}$ \citep{2012ApJ...753...29T}. The initial elemental abundances were $1.55 \times10^{-2}\;(\rm{H})$, $2.17 \times10^{-1}\;(\rm{He})$, $1.13 \times 10^{-4}\;(\rm{C^+})$, $3.32 \times 10^{-5}\;(\rm{N})$, $2.73 \times 10^{-4}\;(\rm{O})$, $3.10 \times 10^{-8}\;(\rm{Si^+})$, and $7.67 \times 10^{-1}\;(\rm{H_2})$. Depletion of species onto dust grains was treated following \citet{1992ApJS...82..167H}, that is, the freeze-out rate ($R_{\rm f}$) is given by
\begin{equation}
R_{\rm f} = S(T) \pi a^2 n_{g^-} v_{\rm th}    
\end{equation}
where $S(T)$ is a temperature-dependent sticking coefficient, $a$ is the grain radius, and $v_\mathrm{th}$ is the thermal velocity of the species under consideration. Given our choices for these free parameters, the depletion timescale at a density of $10^4\,\rm{cm^{-3}}$ is $\sim0.56\,\mathrm{Myr}$, in agreement with simpler estimates by \citet{2007ARA&A..45..339B}. Compared to the dynamical timescale (i.e., the free-fall time for a supercritical cloud), the depletion timescale is 1.5 to 2 times longer. At densities closer to $10^5\,\rm{cm^{-3}}$, on the other hand, the depletion timescale is a factor of 2 shorter. The effect of depletion at these densities can be clearly seen in the abundance maps shown in \citet{TritsisBasuFederrath2025} (i.e., see their fig.~2). For more details on the chemical model we refer the reader to \citet{2016MNRAS.458..789T, 2022MNRAS.510.4420T}.

\subsection{Radiative transfer to generate synthetic observations}
\label{sec:radiation_trans}
The turbulent velocity dispersion in molecular clouds can only be directly measured through radio observations of spectral lines. However, isolating the contribution of turbulence in such observations is challenging due to the complex physical effects that influence the shape, peak location, and width of the spectral lines. To better understand the impact of RT on the measurement of turbulent velocity dispersion in real observations of molecular clouds and cores, we generate synthetic spectral line observations of CO rotational transitions, commonly used in observations. To this end, we make use of the multilevel non-local thermodynamic equilibrium (non-LTE) RT code $\textsc{PyRaTE}$ \citep{2018MNRAS.478.2056T,2024A&A...692A..75T}.

The inputs to \textsc{PyRaTE} from our MHD chemo-dynamical simulation are the $\hyd$ and CO number densities, the temperature in each grid cell, and the three velocity components. Additionally, we use the Einstein and collisional coefficients from the \textsc{LAMBDA} database \citep{2005A&A...432..369S}. The level populations in each rotational level and for every grid point are initially calculated by solving the statistical equilibrium equations (SEEs) under LTE conditions. That is, the escape probabilities of newly created photons are set to unity across all directions, and consequently, only spontaneous emission and collisional excitation and de-excitation are taken into account. Once the level populations are calculated under LTE conditions, we then compute the infinitesimal optical depth inside each grid point. The total optical depth in each direction is then determined by summing the infinitesimal optical depth of all grid points along that LOS. During this process, we only add the grid points whose velocity difference with the grid point under consideration is less than the thermal linewidth. An average optical depth for the grid point under consideration is then computed from all directions with more weight assigned to the directions with lower optical depths. Based on these newly computed optical depths we then calculate new escape probabilities for each pair of rotational levels and solve again the SEEs. As a result, in this new step during this $\Lambda$ iteration, stimulated emission and absorption are also taken into account. Additionally, given our approach for calculating the optical depth where we include variations in velocity, we consider the coupling of level populations throughout the entire cloud. Finally, the process continues until the level populations for all rotational levels simultaneously converge.

To perform our RT simulations we use a spectral resolution of $0.05\,\rm{km\,s^{-1}}$ and 64~points in velocity/frequency space. The contribution from the cosmic-microwave-background (CMB) radiation is also considered in our calculations, and we assume that it is the only external radiation penetrating the cloud. Finally, for the purposes of this study, our spectra have an infinite signal-to-noise ratio and we do not consider any additional instrumental complications. The final output from $\textsc{PyRaTE}$ is a PPV spectral cube.

\subsection{Moment maps}
\label{sec:mom_maps}
While analysing spectral-line observations, moment maps are powerful diagnostic tools as they provide information regarding the integrated intensity, velocity and LOS velocity dispersion of the cloud. Here, we consider a LOS that combines the $x$ and $z$ axes of our MHD simulation as
\begin{align}\label{eqn:los}
    \mathrm{\textbf{LOS}} = \frac{1}{\sqrt{2}} \left(\hat{\textbf{x}} + \hat{\textbf{z}}\right)  &= \frac{1}{\sqrt{2}}\begin{pmatrix}
           1 \\
           0 \\
           1
         \end{pmatrix}.
  \end{align}
Therefore, we effectively ``observe'' the cloud at a 45$^\circ$ angle with respect to the $x$ and $z$ axes. This specific choice is motivated by the anisotropic collapse of the cloud, which occurs predominantly along the $z$ axis due to the presence of the magnetic field. By adopting this projection angle, we enhance gradients in the first-moment (i.e., velocity centroid) map. This occurs because the spectra from regions above and below the cloud's ``midplane'' will exhibit distinct blue- and red-shifted features induced by the gravitational collapse.

In contrast, for LOS that coincide with one of the principal axes of the grid, no prominent large-scale systematic motions would be seen in the first-moment maps. Undoubtedly, the effects of gravity will still manifest to some extent, but these effects will be significantly less pronounced compared to the LOS defined via Eq.~\ref{eqn:los}. As a result, for the remainder of this paper, the LOS defined in Eq.~\ref{eqn:los} will be the primary focus of our analysis. Other moment maps where the LOS aligns with the $x$ and $z$ axes are discussed in Appendix~\ref{app:LOS}. For each LOS, we generate moment maps for the three cases considered here; actual synthetic spectra for CO ($J=1 \rightarrow 0$ and $J=2 \rightarrow 1$) and an ``Ideal'' case.

\subsubsection{Moment maps for the synthetic spectra}
\label{sec:mom_maps_syn_spectra}
The moment maps derived from the CO spectra calculated as described in Sec.~\ref{sec:radiation_trans} represent the most realistic scenario considered here. Generally, the $n\nth$ moment is proportional to $v^n$, where $v$ is velocity. For $n=0,\,1$ and 2, we have the zeroth, first and second moments defined by\footnote{\href{https://spectral-cube.readthedocs.io/en/latest/moments.html}{https://spectral-cube.readthedocs.io/en/latest/moments.html}}
\begin{gather}
    M_0 = \int \omegav dv,\\
    M_1 = M_0^{-1} \int \omegav v dv, \;\mathrm{and}\\
    M_2 = \left(M_0^{-1} \int  \omegav (v-M_1)^2 dv \right)^{1/2},
\label{eqn:mom_0_1_2}
\end{gather}
where $M_\mathrm{n}$ is the $n\nth$ moment and $\omegav$ is a quantity that depends upon whether we consider a CO or Ideal case, as described below. Physically, the zeroth-moment map represents the integrated intensity of the emission line under consideration. It provides information on the cloud’s structure, as projected on the plane of the sky (POS). The first-moment map, or the velocity-centroid map, represents the intensity-weighted average velocity along the LOS. It provides information regarding the bulk motion of the cloud as well as turbulent motions on the POS \citep{2016ApJ...832..143F,stewart_and_federrath}. The second-moment map measures the velocity dispersion along each LOS. 

\subsubsection{The Ideal case}
\label{sec:mom_maps_opt_thin}
For the moment maps calculated as described in the previous section, we simultaneously consider both the effect of chemistry and that of RT. Therefore, there are two layers of complication in order to probe the turbulent velocity dispersion of the cloud. To establish a baseline scenario where such complications are absent, we also produce moment maps under what we term the ``Ideal'' case, defined exactly as in \citet{stewart_and_federrath}. Given that the present study improves upon their work by incorporating chemistry and RT, we require this case as a reference.

To calculate the moment maps for the Ideal case, we replace $\omegav$ in the moment equations of Sec.~\ref{sec:mom_maps_syn_spectra} with the gas density of each computational cell accumulated into the respective velocity channel bin ($v$) that the cell falls into. This means that the effective emission in the Ideal case is assumed to be proportional to the gas density, representing an optically-thin approximation. As such the moment maps are constructed with a weight that is directly proportional to the density of the gas, without RT or chemical effects (which implies that $M_0$ is exactly proportional to the column density). Given this approach, $M_0$ in the Ideal and CO cases differ in units. However, by normalising them to their respective mean values, we can compare the maps directly. $M_1$ and $M_2$ do not require any additional normalisation as they are already normalised by $M_0$, by definition.

\section{Results}
\label{sec:results}
Here we perform a comparative analysis of the moment maps produced from $\cooz$ and $\coto$ observations, and from the ``Ideal'' case (see \S~\ref{sec:mom_maps_opt_thin}). In each figure presented, the corresponding case of each moment map is denoted in the top left corner.
We begin our analysis at a time of one free-fall time ($\tff$), when the central number density of the cloud is $\sim10^4\,\mathrm{cm^{-3}}$. In Sec.~\ref{sec:sim_end} we repeat our analysis at a later time, corresponding to $\ottff$, when the central number density of the cloud is $\sim10^5\,\mathrm{cm^{-3}}$, in order to quantify the impact of the evolutionary stage of the cloud on the results.

\begin{figure*}
\centering\includegraphics[width=0.9\linewidth]{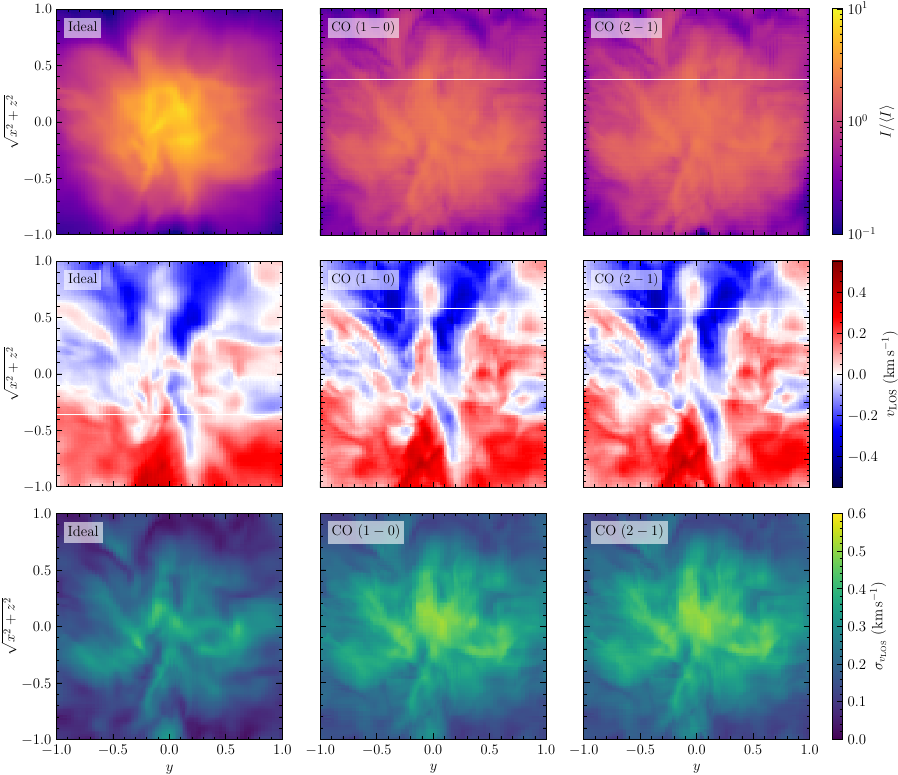}
\caption{Zeroth-moment maps (first row), first-moment maps (second row), and second-moment maps (third row) for the LOS defined via Eq.~\ref{eqn:los} and for $t=\tff$. The left column shows the Ideal (left) case, and the middle and right columns present the CO ($J = 1\rightarrow0$) and ($J = 2\rightarrow1$) transitions, respectively. The zeroth-moment maps (first row) are normalised by their mean values ($\langle I \rangle$, listed in Tab.~\ref{tab:mean_for_zero_mom} of Appendix~\ref{app:mean_for_zero_mom}) to enable direct comparisons between the Ideal and CO cases. The choice of projection angle is motivated by the large gradients in the first-moment (i.e., velocity centroid) map for this projection, i.e., the case for which it is most difficult to isolate turbulent motions \citep{StewartFederrath2022}. We see more of the cloud in the Ideal case than in the CO cases. This is due to optical-depth effects in the CO cases. The gravitational contraction of the cloud is evident from the gradient in the first-moment maps. The first-moment map in the Ideal case is somewhat smoother than in the CO cases. Opposite in nature to the zeroth moment, the second moment shows higher values in the CO cases compared to the Ideal one. The coordinates are in pc.}
\label{fig:mom_map_summary}
\end{figure*}

\subsection{Cloud structure and dynamics revealed by moment maps}
\label{sec:mom_map_summary}

Fig.~\ref{fig:mom_map_summary} presents the three types of moment maps under consideration. The first, second, and third rows correspond to the zeroth-, first-, and second-moment maps, respectively. The columns represent the different cases considered; the first column depicts the Ideal moment maps, the second column shows the $\cooz$ case, and the third column shows the $\coto$ case.

As shown in all three panels in the top row of Fig.~\ref{fig:mom_map_summary}, the zeroth-moment maps of the cloud have higher values in the centre than at the edges. This is due to an increase in $\hyd$ and/or CO number densities in the cloud's centre compared to the outskirts. The increase in the values at the centre is particularly evident for the Ideal case compared to the CO cases. Along with the difference in relative number density of $\hyd$ and/or CO from the centre to the edges of the cloud, this difference between the moment maps of the Ideal and CO cases can also be attributed to self-absorption, and to RT effects. In the Ideal case (upper left panel), the entire cloud, including its far side, is visible, leading to brighter ``emission'' at the centre. In contrast, in the CO maps, the far side of the cloud is obscured, causing a noticeable dip in central brightness. This is discussed in further detail in Appendix~\ref{sec:co_depletion}.

The first-moment maps (middle row of Fig.~\ref{fig:mom_map_summary}), show a pronounced large-scale velocity gradient from the top to the bottom of the maps. This gradient corresponds to the direction of the cloud's collapse. In the Ideal case (left column, middle row), the velocity variations are less pronounced than in the CO maps. This suggests that a slightly larger velocity dispersion would be measured in the CO cases. This can be understood as follows. In the CO case, self absorption results in different velocities being weighted more than others (compared to the Ideal case) while producing the first moment maps. Therefore, such effects lead to more small-scal variations in the velocity centroid in the first-moment maps. We investigate the first-moment maps in more detail in the next sections, as they present a unique way to isolate turbulent motions \citep{2016ApJ...832..143F,stewart_and_federrath}.

The second-moment maps (third row of Fig.~\ref{fig:mom_map_summary}) reveal an opposite trend compared to that observed in the zeroth-moment maps. Specifically, the CO maps exhibit larger velocity dispersions at the centre than at the edges of the maps, compared to the Ideal case. In the Ideal scenario, we are weighting the LOS velocity by the density. Given that the number density changes by $\gtrsim$ three orders of magnitude from the outskirts to the centre of the cloud, we are mostly probing the inner parts of the cloud. In contrast, the CO emission varies less from the outer to the inner parts of the cloud, leading to a more balanced weighting of regions with different velocities. This effect, together with opacity broadening, contribute to the broader velocity distribution.

Overall, the moment maps for the Ideal case and the CO cases exhibit notable visual differences, which are further analysed with spatial variation maps in Appendix~\ref{app:mom_spatial_variations}. It is noteworthy that among all the maps presented, the first-moment maps exhibit the least differences between the CO observations and the Ideal scenario. Hence, we find that the first-moment maps provide the most robust measure of turbulence, and thus, our primary focus will be on the first-moment maps.

\begin{figure*}
\centering\includegraphics[width=0.9\linewidth]{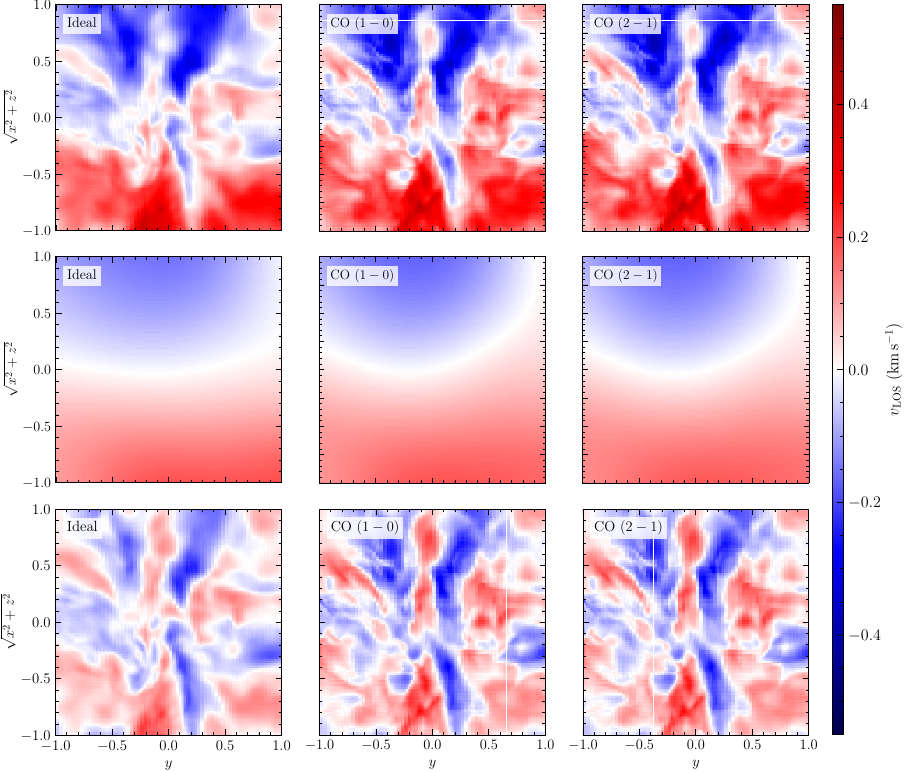}
\caption{The top panels are the same as the middle row of Fig.~\ref{fig:mom_map_summary}, i.e., the original first-moment maps for the three cases considered (from left to right: Ideal, $\cooz$, and $\coto$). These are the starting point of the turbulence-isolation method. First, the original maps are smoothed with a Gaussian low-pass filter (see details in main text), shown the 2nd row, which represent the large-scale systematic contributions (in-fall, rotation, or shear). Subtracting the smoothed fields from the original ones yields the turbulence-isolated maps, shown in the bottom panels, which are used to quantify the turbulent velocity dispersion.}
\label{fig:turb_isolation}
\end{figure*} 

\subsection{Isolating turbulence in the first-moment maps}
\label{sec:turb_isolation}
Turbulence is not the sole type of motion within clouds. Systematic motions, including large-scale rotation, gravitational collapse, and shear, are commonly observed in the first-moment maps of molecular clouds. It is crucial, however, to differentiate these large-scale motions from turbulence. Previous studies, including \citet{2016ApJ...832..143F}, \citet{2018MNRAS.477.4380S, 2019MNRAS.487.4305S}, and \citet{2021MNRAS.500.1721M}, have employed gradient subtraction techniques to isolate turbulence, often using linear gradients. In this paper, we adopt a different approach by applying Gaussian smoothing to the first-moment maps to remove low-mode (i.e., low-Fourier-mode) contributions (see also Appendix~\ref{app:spect}), in line with the method first used in \citet{2024MNRAS.530.4317G}. Our choice of the kernel size for the Gaussian filtering process is at $k=2$, which is a FWHM of $L/2$, with $L$ being the diameter of the cloud. This choice is motivated by the fact that the largest turbulent modes exist at $k=2$ \citep{2016ApJ...832..143F}.

The first and last rows of Fig.~\ref{fig:turb_isolation} compare the first-moment maps before (first row) and after (last row) turbulence isolation, respectively. The middle row shows the result of Gaussian smoothing of the original first-moment maps. The smoothed maps are subtracted from the original ones, which results in the turbulence-isolated maps. The columns represent the different cases considered, as in Fig.~\ref{fig:mom_map_summary}. The absence of large-scale gradients in the last row of Fig.~\ref{fig:turb_isolation} indicates that most of the large-scale systematic motions have been removed, and turbulent motions have been isolated.

Comparing the maps after turbulence isolation (last row of Fig.~\ref{fig:turb_isolation}), we find that the velocity variations in the Ideal case are somewhat smaller than in the CO cases, due to the differences in small-scale structure between the Ideal and CO cases. Comparing the second and third columns, i.e., $\cooz$ vs.~$\coto$, we find that the small-scale features and velocity ranges are similar, i.e., there are no significant differences between the $\cooz$ and $\coto$ lines. We now quantify these differences via probability density functions (PDFs).

\begin{figure}
\centering\includegraphics[width=0.9\linewidth]{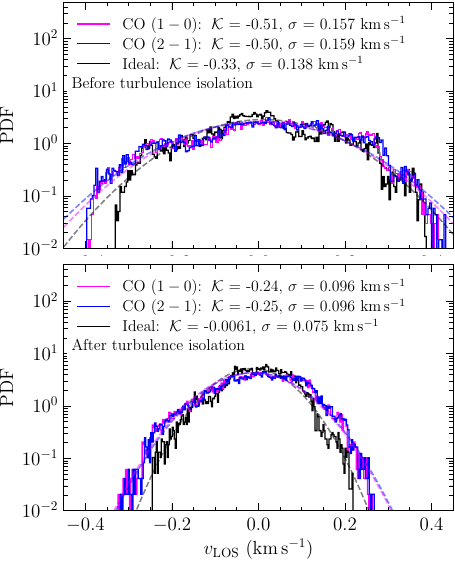}
\caption{PDFs of the first-moment maps shown in Fig.~\ref{fig:turb_isolation} before (top panel) and after (bottom panel) turbulence isolation, for the Ideal, $\cooz$, and $\coto$ cases. Gaussian fits for each of the three cases considered are shown as dashed lines for comparison. Before turbulence isolation, the PDFs are broad and have significant non-Gaussian features, as quantified by the standard deviation ($\sigma$) and kurtosis ($\mathcal{K}$), respectively, listed in the legends. After turbulence isolation, the PDFs are closer to Gaussian distributions with significantly smaller standard deviation, as expected after the removal of non-turbulent contributions. Comparing the Ideal and CO cases, we find that the CO cases have somewhat larger $\sigma$ than the Ideal case, which is quantified further and corrected for in Sec.~\ref{sec:effect_on_turb} below.}
\label{fig:pdfs}
\end{figure}

\begin{table*}
\def\arraystretch{1}
\setlength{\tabcolsep}{6pt}
\caption{Standard deviations of the first-moment maps, $\sigvod$, for the Ideal, $\cooz$, and $\coto$ cases, before and after turbulence isolation, for different lines of sight (LOS).}
\makebox[\textwidth][c]{
    \begin{tabular}{ccccccc}
\toprule
    & \multicolumn{2}{c}{Ideal} & \multicolumn{2}{c}{$\cooz$} & \multicolumn{2}{c}{$\coto$} \\
\cmidrule(lr){2-3}\cmidrule(lr){4-5}\cmidrule(lr){6-7}
& \multicolumn{1}{c}{Before} & \multicolumn{1}{c}{After} & \multicolumn{1}{c}{Before} & \multicolumn{1}{c}{After} & \multicolumn{1}{c}{Before} & \multicolumn{1}{c}{After}\\
\cmidrule(lr){2-2}\cmidrule(lr){3-3}\cmidrule(lr){4-4}\cmidrule(lr){5-5}\cmidrule(lr){6-6}\cmidrule(lr){7-7}
LOS & $\sigvod \left[\,\kmps\right]$ & $\sigvod \left[\,\kmps\right]$ & $\sigvod \left[\,\kmps\right]$ & $\sigvod \left[\,\kmps\right]$ & $\sigvod \left[\,\kmps\right]$ & $\sigvod \left[\,\kmps\right]$ \\ \midrule
                      (0 0 1) &  0.080$^{+0.005}_{-0.005}$ &  0.068$^{+0.006}_{-0.006}$ &  0.087$^{+0.006}_{-0.007}$ &  0.075$^{+0.005}_{-0.005}$ &  0.088$^{+0.006}_{-0.007}$ &  0.075$^{+0.004}_{-0.005}$ \\
 $\frac{1}{\sqrt{2}}$ (1 0 1) &  0.138$^{+0.009}_{-0.008}$ &  0.075$^{+0.005}_{-0.006}$ &  0.157$^{+0.010}_{-0.011}$ &  0.096$^{+0.006}_{-0.007}$ &  0.159$^{+0.010}_{-0.008}$ &  0.096$^{+0.006}_{-0.006}$ \\
                      (1 0 0) &  0.076$^{+0.007}_{-0.005}$ &  0.069$^{+0.005}_{-0.004}$ &  0.097$^{+0.006}_{-0.006}$ &  0.088$^{+0.006}_{-0.007}$ &  0.098$^{+0.008}_{-0.008}$ &  0.088$^{+0.006}_{-0.006}$ \\
\bottomrule
\end{tabular}
}
\label{tab:sigv}
\end{table*}

Fig.~\ref{fig:pdfs} shows the first-moment PDFs before (top panel) and after (bottom panel) turbulence isolation. The different colours represent the three cases under consideration. The values of $\sigvod$ before turbulence isolation are larger than those after turbulence isolation by 35-50\%. This is due to the cloud's gravitational in-fall, which broadens the range and dispersion of velocities. After turbulence isolation, the PDFs are closer to Gaussian distributions, consistent with the expected velocity PDF of turbulent flows \citep{2013MNRAS.436.1245F}. However, the PDFs post-turbulence isolation still retain certain non-turbulent/non-Gaussian characteristics, primarily attributable to the incomplete separation of turbulent components and the presence of intermittent events in the turbulent flow \citep{1995A&A...293..840F, 2008A&A...481..367H, 2010A&A...512A..81F,2016ApJ...832..143F}. The skewness and kurtosis can serve as quantitative indicators of the deviation from a Gaussian distribution. The PDFs in Fig.~\ref{fig:pdfs} have hardly any asymmetry before and after turbulence isolation, however, the kurtosis values labelled in the figure legends are closer to zero (Gaussian) after turbulence isolation, as expected.

To better understand how turbulence isolation depends on the different LOS, Tab.~\ref{tab:sigv} lists the $\sigvod$ values for the three cases considered, along three different LOS, before and after turbulence isolation. Before turbulence isolation, $\sigvod$ is largest along our main LOS (c.f., Eq.~\ref{eqn:los}) due to the large gradient induced by gravitational collapse in the first moment. However, after turbulence isolation, the values become nearly uniform across all LOS, as intended \citep[see also][]{stewart_and_federrath}. Additionally, the CO cases show a larger velocity dispersion than the Ideal cases (as qualitatively mentioned in Sec.~\ref{sec:mom_map_summary}), which is quantified further below. This fact also persists for other viewing angles.

\subsection{Influence of time evolution}
\label{sec:sim_end}
In this section, we investigate how turbulence evolves over time and how changes in the chemical evolution of the cloud can further influence the way in which the turbulence is measured.

\begin{figure*}
\centering\includegraphics[width=0.9\linewidth]{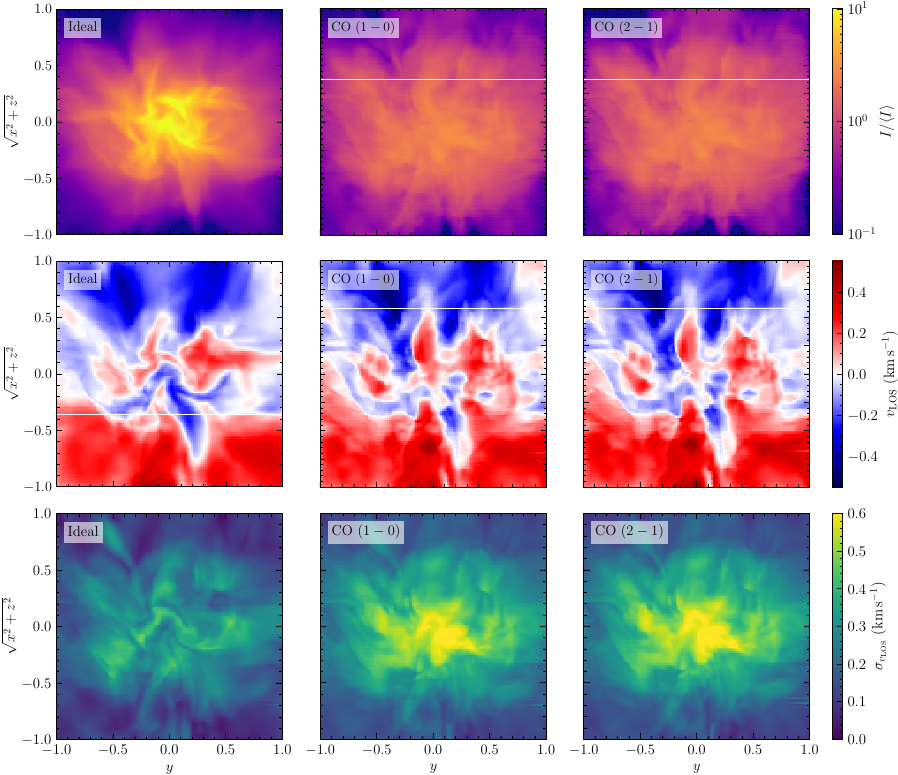}
\caption{Same as Fig.~\ref{fig:mom_map_summary}, but at $t=\ottff$. Trends observed in this figure are similar to those in Fig.~\ref{fig:mom_map_summary}.}
\label{fig:sim_end_mom_map_sum}
\end{figure*}

Fig.~\ref{fig:sim_end_mom_map_sum} shows the same as Fig.~\ref{fig:mom_map_summary}, but at $\ottff$. We see that the general trends observed are similar to the earlier time, with some differences. In the zeroth-moment maps (first row), the maximum value of the relative intensity for the three maps is, on average, larger than that in Fig.~\ref{fig:mom_map_summary} by $\approx 60\%$. A similar trend is seen in the first-moment maps (second row), where the value of the maximum velocity is higher than that in Fig.~\ref{fig:mom_map_summary}, on average, by $\approx23\%$, and for the second-moment maps (third row) the value of the maximum dispersion is $\approx 18\%$ larger. As the cloud collapses further, CO progressively depletes onto dust grains in the densest regions (e.g., fig.~2 in \citealt{TritsisBasuFederrath2025}). This leads to a further reduction in the central emission for the CO cases in comparison to the Ideal case. Although there are quantitative differences in the moment maps of Fig.~\ref{fig:sim_end_mom_map_sum}, the qualitative differences between the three cases remain the same over time. 

\begin{figure*}
\centering\includegraphics[width=0.9\linewidth]{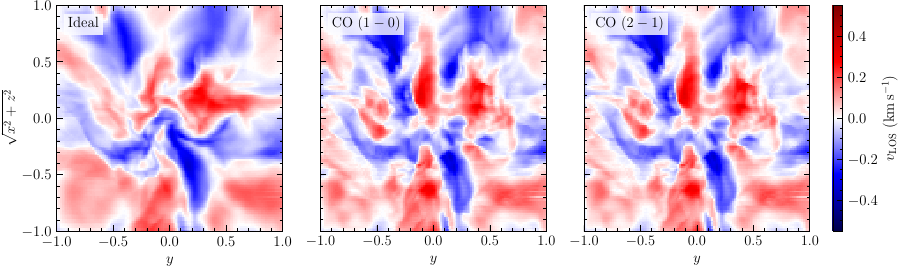}
\caption{Same as Fig.~\ref{fig:turb_isolation} bottom panels, but at $t=\ottff$.}
\label{fig:sim_end_turb_iso}
\end{figure*}

\begin{figure}
\centering\includegraphics[width=0.9\linewidth]{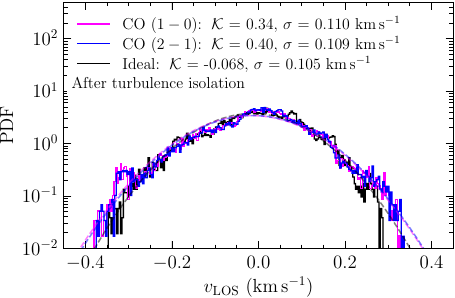}
\caption{Same as Fig.~\ref{fig:pdfs} bottom panel, but at $t=\ottff$. The general features of the turbulence isolation remain valid at $\ottff$, with the standard deviation having increased by $\sim15-30\%$ at the later evolutionary stage shown here compared to $t=\tff$, due to additional turbulence being driven by the collapse of the cloud.}
\label{fig:sim_end_pdfs}
\end{figure}

Fig.~\ref{fig:sim_end_turb_iso} shows the effect of turbulence isolation, as in the bottom panels of Fig.~\ref{fig:turb_isolation}, but at $\ottff$. As observed, the CO cases (middle and right) exhibit a larger range in velocities compared to the Ideal case (left), which we quantify in Fig.~\ref{fig:sim_end_pdfs}, showing the respective PDFs. After turbulence isolation, $\sigvod$ for the Ideal case is smaller than that of the CO cases by about 10\%. This is a smaller change than that in Fig.~\ref{fig:pdfs}, where the decrease was approximately 20\%. Overall, the turbulent velocity dispersion has increased by 15-30\% at $\ottff$ compared to $t=\tff$, due to turbulence being replenished and added by the gravitational collapse of the cloud.

In summary, we find that the general trends and relative differences between the CO cases and the Ideal case remain similar at $\ottff$ when the cloud has contracted substantially more compared to $t=\tff$.

\section{The effects of CO chemistry and optical depth on turbulence reconstruction}
\label{sec:effect_on_turb}

\subsection{Correcting for chemical and optical-depths effects in CO line observations}

\begin{table*}
\def\arraystretch{1.1}
\setlength{\tabcolsep}{10pt}
\caption{Turbulent velocity dispersion measured from the first moment, $\sigvod$, and CO-to-Ideal correction factor, $\rco$ (defined in Eq.~\ref{eq:rco}), after turbulence isolation the Ideal, $\cooz$, and $\coto$ cases, for different LOS, and for $t=\otff$ and $\ottff$. The averages over the LOS are listed in the last row of the respective time segment ($t=\otff$ and $\ottff$), and the averages over all LOS and over the two times are listed in the very last row. The errors are obtained by bootstrapping.}
\makebox[\textwidth][c]{
    \begin{tabular}{cccccc}
\toprule
    & \multicolumn{1}{c}{Ideal} & \multicolumn{2}{c}{$\cooz$} & \multicolumn{2}{c}{$\coto$} \\
    \cmidrule(lr){2-2}\cmidrule(lr){3-4}\cmidrule(lr){5-6} 
LOS & $\sigvod \left[\,\kmps\right]$ & $\sigvod \left[\,\kmps\right]$ & $\rcooz$ & $\sigvod \left[\,\kmps\right]$ & $\rcoto$ \\ \midrule
\multicolumn{6}{c}{$t=\otff$} \\
\midrule
                      (0 0 1) &  0.068$^{+0.006}_{-0.006}$ &  0.075$^{+0.005}_{-0.005}$ &  0.91$^{+0.08}_{-0.08}$ &  0.075$^{+0.004}_{-0.005}$ &  0.91$^{+0.08}_{-0.08}$ \\
 $\frac{1}{\sqrt{2}}$ (1 0 1) &  0.075$^{+0.005}_{-0.006}$ &  0.096$^{+0.006}_{-0.007}$ &  0.79$^{+0.09}_{-0.06}$ &  0.096$^{+0.006}_{-0.006}$ &  0.79$^{+0.08}_{-0.07}$ \\
                      (1 0 0) &  0.069$^{+0.005}_{-0.004}$ &  0.088$^{+0.006}_{-0.007}$ &  0.79$^{+0.08}_{-0.07}$ &  0.088$^{+0.006}_{-0.006}$ &  0.79$^{+0.09}_{-0.08}$ \\
\midrule
                 LOS Averages &  0.071$^{+0.005}_{-0.005}$ &  0.087$^{+0.005}_{-0.006}$ &  0.83$^{+0.08}_{-0.07}$ &  0.086$^{+0.006}_{-0.006}$ &  0.83$^{+0.08}_{-0.08}$ \\
\midrule
\multicolumn{6}{c}{$t = \ottff$} \\
\midrule
                      (0 0 1) &  0.082$^{+0.006}_{-0.006}$ &  0.075$^{+0.006}_{-0.006}$ &  1.08$^{+0.12}_{-0.08}$ &  0.076$^{+0.007}_{-0.007}$ &  1.08$^{+0.16}_{-0.13}$ \\
 $\frac{1}{\sqrt{2}}$ (1 0 1) &  0.105$^{+0.007}_{-0.007}$ &  0.110$^{+0.010}_{-0.008}$ &  0.96$^{+0.09}_{-0.10}$ &  0.109$^{+0.008}_{-0.008}$ &  0.96$^{+0.10}_{-0.09}$ \\
                      (1 0 0) &  0.070$^{+0.005}_{-0.004}$ &  0.093$^{+0.007}_{-0.006}$ &  0.75$^{+0.07}_{-0.08}$ &  0.093$^{+0.007}_{-0.006}$ &  0.74$^{+0.09}_{-0.07}$ \\
\midrule
                 LOS Averages &  0.086$^{+0.006}_{-0.006}$ &  0.093$^{+0.007}_{-0.007}$ &  0.93$^{+0.09}_{-0.08}$ &  0.093$^{+0.007}_{-0.007}$ &  0.93$^{+0.12}_{-0.09}$ \\
\midrule
 Time average of LOS averages &  0.078$^{+0.005}_{-0.006}$ &  0.090$^{+0.006}_{-0.007}$ &  0.88$^{+0.09}_{-0.08}$ &  0.089$^{+0.006}_{-0.006}$ &  0.88$^{+0.10}_{-0.09}$ \\
\bottomrule
\end{tabular}
}
\label{tab:rco}
\end{table*}

Here we summarise the results of the 1D turbulent velocity dispersion, $\sigvod$, from the turbulence-isolated first-moment maps, for the Ideal, $\cooz$, and $\coto$ cases, for different LOS, and for $t=\otff$ and $\ottff$. All relevant values are listed in Tab.~\ref{tab:rco} as well as their LOS and time averages.

Thus, we now have everything in place to quantify how the $\sigvod$ values for the CO cases differ from their Ideal counterparts by defining a CO-to-Ideal conversion factor, $\rco$, as
\begin{equation}\label{eq:rco}
    \rco = \frac{\sigvod\,\text{(Ideal)}}{\sigvod\,(\mathrm{CO})},
\end{equation}
where the ``CO'' placeholder can be either `$\mathrm{CO,1\!-\!0}$' referring to $\cooz$ or `$\mathrm{CO,2\!-\!1}$' referring to $\coto$. The $\rco$ values are listed in Tab.~\ref{tab:rco}. We see that they are usually $<1$, with the average values over all LOS and the two simulation times being $\rcooz=0.88^{+0.09}_{-0.08}$ and $\rcoto=0.88^{+0.10}_{-0.08}$. Thus, the CO lines tend to overestimate the Ideal $\sigvod$ by $1-\rco^{-1}\sim12-14\%$ on average, with a 1-sigma maximum overestimation of $41\%$.

\citet{stewart_and_federrath} used simulations corresponding to the Ideal case discussed in this paper, in order to provide a method to reconstruct the 3D turbulent velocity dispersion from $\sigvod$, given by
\begin{equation}\label{eq:original_correction_factor}
    \begin{split}
        \sigvtd & = \csf \times \sigvod\,\text{(Ideal)} \\
                & = \csf \times \rco \times \sigvod\,(\mathrm{CO}),
    \end{split}
\end{equation}
which involves another conversion factor, $\csf$, as listed in \citet[][tab.~E1]{stewart_and_federrath}. In the 2nd line of this equation, we have inserted Eq.~(\ref{eq:rco}), such that a user measuring $\sigvod$ from the turbulence-isolated first-moment map obtained with CO spectral observations, can directly estimate the 3D turbulent velocity dispersion ($\sigvtd$), by multiplying both the conversion factor $\csf$ \citep[provided by][]{stewart_and_federrath} and the CO-to-Ideal correction factor $\rco$ obtained in this work and listed in Tab.~\ref{tab:rco}.

\subsection{Implications for previous 3D turbulence estimates from CO line observations}

\citet{GerrardEtAl2023,2024MNRAS.530.4317G} compile a variety of sources from the literature measuring the density dispersion–-Mach number relation in different environments. This relation is important as it provides an estimate of the turbulence driving mode in different regions of the interstellar medium. Crucially, it requires an estimate of the 3D turbulent Mach number, which in turn requires an estimate of the 3D turbulence velocity dispersion, $\sigvtd$. Many measurements of this in molecular clouds and cores rely on $^{12}\co$ and/or $^{13}\co$ line observations, including the studies by \citet{1997ApJ...474..730P}, \citet{Brunt2010}, \citet{GinsburgFederrathDarling2013}, \citet{KainulainenTan2013}, \citet{2021MNRAS.500.1721M}, and \citet{ShardaEtAl2022}. 

If a correction factor, as discussed above, were to be applied to these previous observational estimates, the value of $\mach\propto\sigvtd$ obtained based on their works, would be reduced by the $\rco$ factor obtained here. This in turn would affect some of the estimates of the turbulence driving parameter, $b\propto\mach^{-1}\propto\sigvtd^{-1}$ obtained in \citet{Brunt2010}, \citet{2021MNRAS.500.1721M}, and \citet{ShardaEtAl2022}, such that their obtained values would increase by a factor $\rco^{-1}$. Considering that $\rco\sim0.88^{+0.09}_{-0.08}$ on average (c.f., Tab.~\ref{tab:rco}), the correction is relatively small, but can nevertheless play a role when high precision is required and the velocity dispersion is used for further processing (e.g., in theoretical models predicting star formation activity, etc). However, the value of $\rco$ is not always lower than unity. In some cases in Tab.~\ref{tab:rco} we see that $\rco \gtrsim 1$ when the cloud is observed along its collapse/rotation axis at a later evolutionary stage where the collapse may have had some influence and CO gets depleted. These cases are discussed in detail in Appendix~\ref{sec:rco_gret_1}. With better data and therefore smaller uncertainties expected from future observations, the $\rco$ correction may become more critical.

\subsection{Caveats}

\subsubsection{Generalising $\rco$}

We consider the effects of chemistry and radiation transfer when using CO lines to probe the turbulent velocity dispersion. A single collapsing cloud in two different timesteps was studied, providing an $\rco$ correction factor (Eq.~\ref{eq:rco}) relative to the previously studied Ideal case. However, a large body of literature \citep{2015MNRAS.450.4424B,2015MNRAS.452.2057C,2019MNRAS.486.4622C,2020ApJ...903..142G,2021MNRAS.502.2701B,2021ApJ...920...44H,2022ApJ...931...28H,2024MNRAS.527.8886B} has shown that the CO emission depends on the environmental conditions of the ISM and as a result it can trace different density distributions. Variations in the external radiation field, cosmic-ray ionisation rate, and/or in the metallicity of the cloud will affect the thermal balance (and therefore the gas temperature), the CO abundance, and consequently the CO level populations, optical depth, and resulting emission. Thus, the $\rco$ correction factors provided here strictly only apply to the conditions studied in the simulations, namely those of the typical ISM in the solar neighbourhood. However, given that $\rco$ is found to be less than unity for both simulation timesteps (where the CO abundance distributions are very different) and for most projection angles, implies that turbulent motions, as probed by CO observations, are likely overestimated compared to the Ideal case.

\subsubsection{Probing different lines of CO}
The purpose of this study is to focus specifically on the effects of CO lines on the measurement of turbulence, and to enhance our understanding of different excitation and radiation transfer effects. The reason for focussing on the $\cooz$ and $\coto$ lines is that CO is one of the most widely used molecules to probe the kinematics of turbulent molecular clouds. While the $\cooz$ line is stronger and well studied, it can become optically thick in denser environments. On the other hand, we use the $\coto$ line as a benchmark to understand how radiative transfer affects the measurements, i.e., since both lines should be unaffected by chemistry (i.e., probing exactly the same regions of the cloud) with the only difference being excitation and radiation transfer properties. Although we find that the two lines are quite similar, there are some differences. The $\coto$ line is optically thinner than the $\cooz$ line, so it probes deeper into the cloud. Its critical density and excitation temperature are ten and three times larger than the $\cooz$ line, respectively \citep{Shirley_2015}. Different tracers also have different excitation temperatures and critical densities. Using higher-density tracers one can probe other physical regimes compared to that of CO. In future work, the effect of other molecules, like HCN or HCO$^+$ may be studied to better understand the specifics of each of the molecular lines when used for kinematic inference, in particular turbulence measurements.

\section{Conclusions}
\label{sec:conclusions}

In this paper, we have quantified the effects of chemistry, excitation, and radiation transfer (RT) in spectral line observations using the $\cooz$ and $\coto$ lines, by creating synthetic observations of an MHD simulation of a collapsing, turbulent cloud. A particular focus is the first-moment map, which can be used to estimate the turbulent velocity fluctuations of a molecular cloud. Using a turbulence-isolation method, which removes a smooth version of the first-moment map ($M_1$), representing non-turbulent contributions from the original $M_1$, the standard deviation of $M_1$ provides the turbulent 1D velocity dispersion, $\sigvod$. However, we find that using the CO lines to obtain $M_1$ leads to an overestimate of $\sigvod$ compared to the optically-thin (``Ideal'') case, which we quantified in detail in this work. Our main conclusions are as follows:
\begin{enumerate}
    \item The moment maps for the Ideal and CO cases show distinct visual differences (c.f., Fig.~\ref{fig:mom_map_summary}), with the zeroth- and second-moment maps showing co-spatial differences by factors of up to a few between the Ideal and CO cases. The $M_1$ is the least affected and provides a good basis for turbulent velocity reconstruction.
    \item Performing turbulence isolation (Sec.~\ref{sec:turb_isolation}) on $M_1$ (c.f., Figs.~\ref{fig:turb_isolation} and~\ref{fig:pdfs}), we can successfully remove large-scale, non-turbulent contributions in $M_1$. This method works similarly well in the Ideal and CO cases.
    \item The basic methods work regardless of the evolutionary stage or line of sight towards the cloud, as quantified in Tab.~\ref{tab:rco}, which lists the 1D turbulent velocity dispersions of all cases considered.
    \item Based on these measurements, we can construct a CO-to-Ideal correction factor, $\rco$ (Eq.~\ref{eq:rco}). While there is some variation of $\rco$ for different LOS and evolutionary stages of the cloud, the time- and LOS-averaged values provide a reasonable average correction, $\rcooz=0.88^{+0.09}_{-0.08}$ and $\rcoto=0.88^{+0.10}_{-0.08}$. However, this value is for typical ISM conditions in the solar neighbourhood and may deviate into the $\rco \gtrsim1$ regime for later times when CO depletes in the dense gas, and only when the cloud is observed along the LOS along which the cloud collapses.
\end{enumerate}

Thus, previous measurements of the turbulent velocity dispersion based on CO lines may have been overestimated by $10-15\%$ on average, with potential maximum 1-sigma overestimates by as much as $40\%$, based on this work.

\section*{Acknowledgements}
We thank the anonymous referee for their comments, which helped us improve this work. C.~F.~acknowledges funding provided by the Australian Research Council (Discovery Projects DP230102280 and DP250101526), and the Australia-Germany Joint Research Cooperation Scheme (UA-DAAD). A.~T.~acknowledges support by the Ambizione grant no. PZ00P2\_202199 of the Swiss National Science Foundation (SNSF). We further acknowledge high-performance computing resources provided by the Leibniz Rechenzentrum and the Gauss Centre for Supercomputing (grants~pr32lo, pr48pi and GCS Large-scale project~10391), the Australian National Computational Infrastructure (grant~ek9) and the Pawsey Supercomputing Centre (project~pawsey0810) in the framework of the National Computational Merit Allocation Scheme and the ANU Merit Allocation Scheme. This research was enabled in part by support provided by SHARCNET (Shared Hierarchical Academic Research Computing Network) and Compute/Calcul Canada and the Digital Research Alliance of Canada. The simulation software, \texttt{FLASH}, was in part developed by the Flash Centre for Computational Science the University of Chicago and the Department of Physics and Astronomy of the University of Rochester. Jayashree Narayan thanks the \href{https://science.anu.edu.au/study/scholarships/future-research-talent-awards-india}{ANU FRT panel} for awarding her with an FRT scholarship, which funded her stay at the Australian National University.

\section*{Data Availability}
The codes and results from the present study will be shared on reasonable request to the corresponding author.



\bibliographystyle{mnras}
\bibliography{main,federrath} 



\appendix

\section{Additional lines-of-sight (LOS)}
\label{app:LOS}
\begin{figure*}
\centering\includegraphics[width=0.9\linewidth]{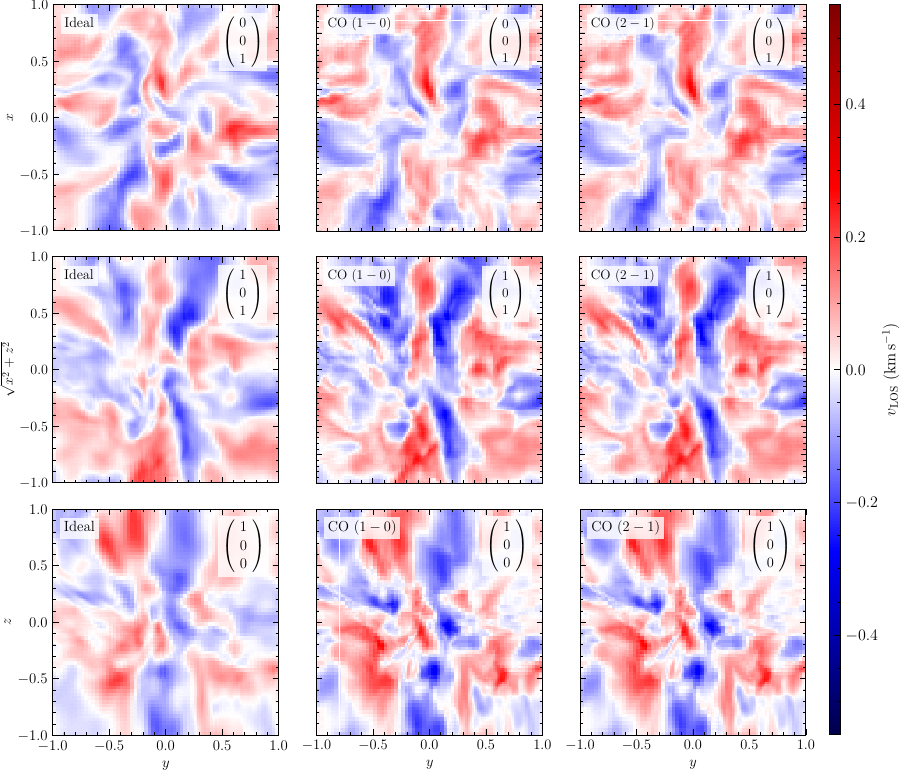}
\caption{Same as Fig.~\ref{fig:turb_isolation} bottom panels, but for different LOS, for $\otff$.}
\label{fig:los_early_time}
\end{figure*}

Fig.~\ref{fig:los_early_time} shows the first-moment maps after turbulence isolation for different projection angles at $\otff$. The top row is for a LOS along the $z$-axis, the middle row is for the LOS shown by Eq.~(\ref{eqn:los}), and the bottom row is for the LOS is along the $x$-axis. The left column shows the Ideal case, the middle column shows the $\cooz$ case and the right column shows the $\coto$ case. Overall, the turbulence isolation performs well for any projection angle \citep[see also][]{stewart_and_federrath}.

\section{Normalisation factors in zeroth moment maps}
\label{app:mean_for_zero_mom}
In Figures~\ref{fig:mom_map_summary} and~\ref{fig:sim_end_mom_map_sum} we normalise the zeroth moment maps by their mean values, $\langle I \rangle$, which are listed in Tab.~\ref{tab:mean_for_zero_mom}. The Ideal case values are displayed in units of column density (cm$^{-2}$) and the CO values are shown in units of brightness temperature, given by
\begin{equation}
    T = \frac{I \times c^2}{2\times \nu^2 \times k_\mathrm{B}},
\label{eqn:brightness_temp}
\end{equation}
where $I$ is the intensity, $c$ is the speed of light, $\nu$ is the frequency of the CO transition under consideration, and $k_\mathrm{B}$ is the Boltzmann constant.

\begin{table}
\caption{Values of $\langle I \rangle$ for the zeroth moment maps in Fig.~\ref{fig:mom_map_summary}.}
\label{tab:mean_for_zero_mom}
\begin{tabular}{cccc}
\hline
Time & Ideal & $\cooz$ & $\coto$ \\ \hline
$\otff$ & $1.8\times10^{21}\,\mathrm{cm}^{-2}$ & $4.4\,\mathrm{K}$ & $7.9\,\mathrm{K}$  \\
$\ottff$ & $1.8\times10^{21}\,\mathrm{cm}^{-2}$  & $4.5\,\mathrm{K}$ & $7.9\,\mathrm{K}$  \\
 \hline
\end{tabular}
\end{table}

\section{Spatial Variation}
\label{app:mom_spatial_variations}
\begin{figure*}
\centering\includegraphics[width=0.66\linewidth]{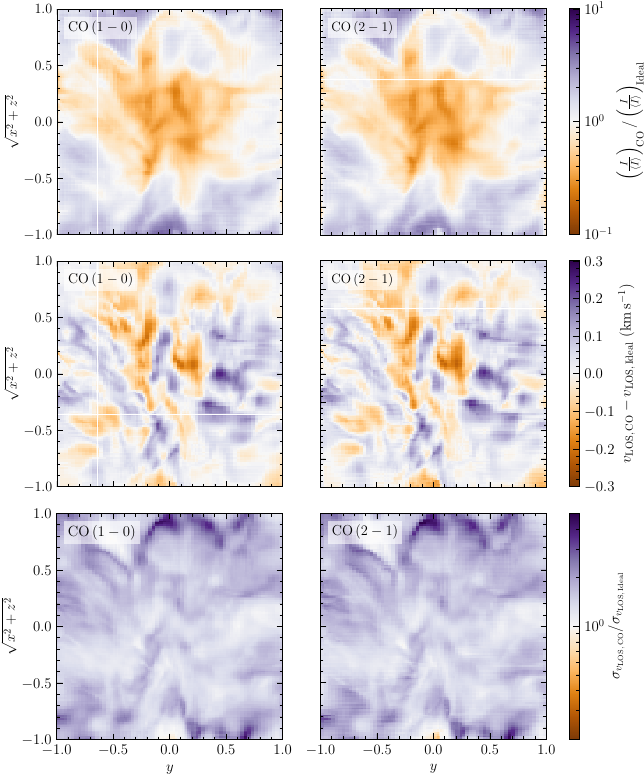}
\caption{Spatial variations between the CO cases and the Ideal case for the zeroth, first, and second moment (from top to bottom).}
\label{fig:mom_spatial_variations}
\end{figure*}

We can quantify the spatial variation in the differences between the CO cases and the Ideal case by plotting ratios and differences as labelled on the colour bars in Fig.~\ref{fig:mom_spatial_variations}. The zeroth- and second-moment maps show variations by factors of several depending on the spatial region of the cloud. The zeroth moment (top panels) is underestimated by CO in the centre and overestimated in the outskirts primarily due to optical-depth effects. The second moment (bottom panels) is generally overestimated by factors of $\sim2-3$. The first moment (middle panels) does not show a strong and/or systematic over-/underestimate with CO, but local differences between the CO cases and the Ideal case can be as high as $\pm\,0.3\,\kmps$.

\section{Spectral analysis}
\label{app:spect}
\begin{figure}
\centering\includegraphics[width=0.9\linewidth]{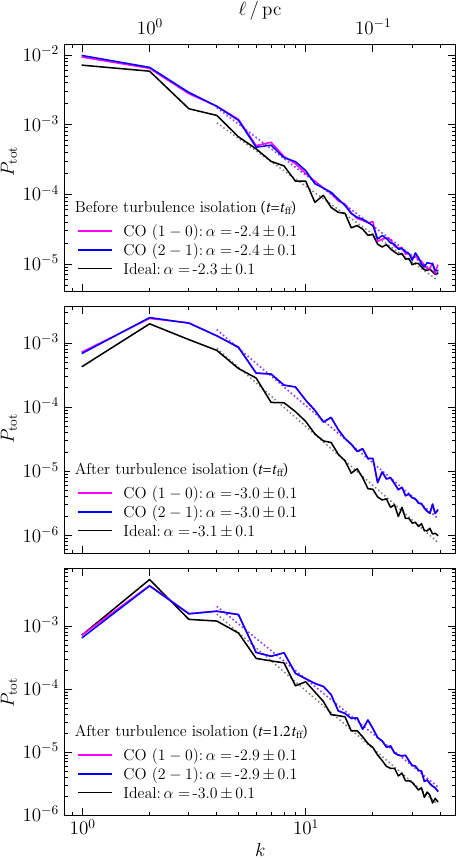}
\caption{Power spectra of the first-moment maps before turbulence isolation (top panel) and after turbulence isolation (middle panel) for the Ideal, $\cooz$ and $\coto$ cases at $t=\tff$. The bottom panel is the same as the middle panel, but at $t=\ottff$. Power-law fits are shown as the dotted lines, with the power-law exponents listed in the respective panel legends.}
\label{fig:spect}
\end{figure}

Fig.~\ref{fig:spect} shows the power spectra of the first-moment maps before (top panel) and after turbulence isolation (middle panel) at $t=\tff$. As expected, the $M_1$ power spectra exhibit larger values before than after turbulence isolation. The $\cooz$ and $\coto$ cases are similar in $k\sim2-10$. Our choice for the size of the Gaussian kernel/scale to remove large-scale non-turbulent velocity contributions is also evident in the power spectra by the drop in power at $k=1$ after turbulence isolation. The spectral shapes of all three cases have become more similar after turbulence isolation, while before, we see differences in the Ideal case compared to the CO cases, especially at high $k$. The spectral slopes after turbulence isolation decrease from $\sim -2.4$ to $\sim -3.0$ across all cases. The dependence on time or evolutionary stage of the cloud is minor (see bottom panel of Fig.~\ref{fig:spect}, which shows the same as the middle panel, but for $t=\ottff$).

\section{Cases with $\rco>1$}
\label{sec:rco_gret_1}
\begin{figure}
\centering\includegraphics[width=0.9\linewidth]{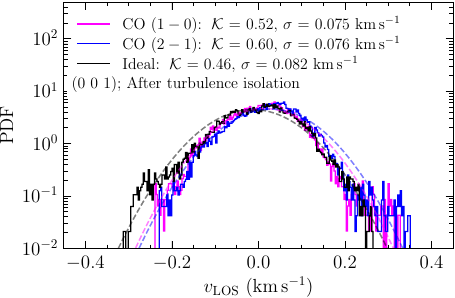}
\caption{Same as Fig.~\ref{fig:pdfs} bottom panel, but for the LOS along the $z$-axis, denoted by (0 0 1), and at $t=\ottff$. Non-Gaussian features induced by the gravitational collapse along this LOS are primarily visible in the Ideal case, while CO has depleted onto grains in the regions most affected by this, and therefore, CO shows less prominent non-Gaussian components. As a result, $\rco>1$, for this particular LOS and towards late times in the collapse.}
\label{fig:sim_end_pdfs_001}
\end{figure}

\begin{figure*}
\centering\includegraphics[width=0.9\linewidth]{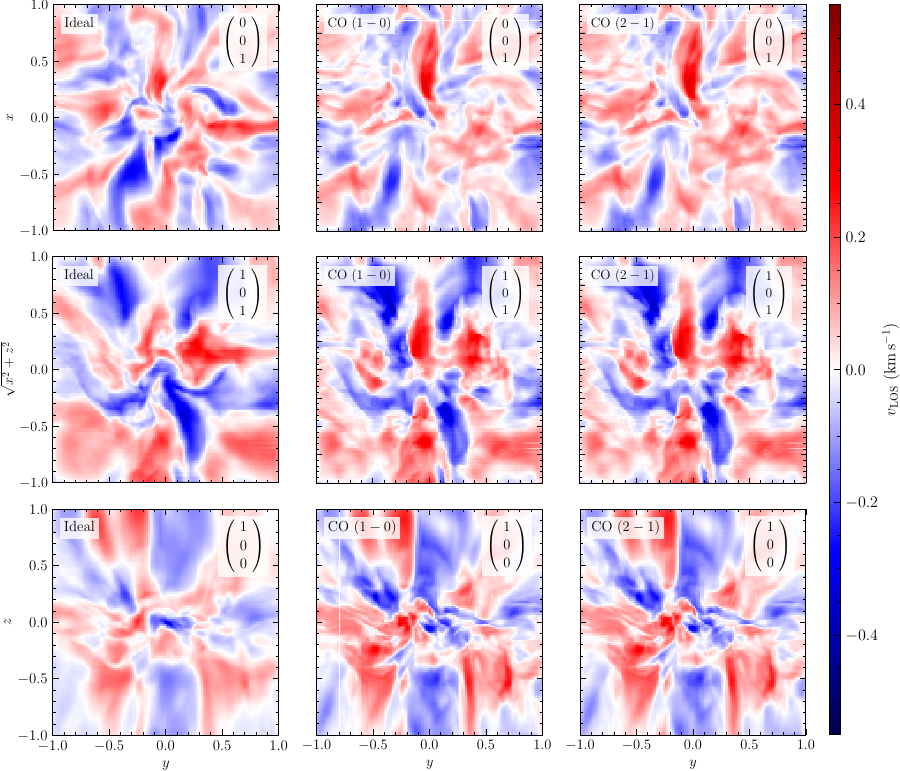}
\caption{Same as Fig.~\ref{fig:los_early_time}, but for the later time during the collapse (at $\ottff$).}
\label{fig:los_late_time}
\end{figure*}

In Table~\ref{tab:rco}, we see that for the LOS $(0\,0\,1)$ case at $t=\ottff$, the CO-to-Ideal conversion factor $\rco>1$. The $(0\,0\,1)$ LOS is the $z$-axis, which is also the axis where the effect of gravity is more pronounced at later times. We can quantify this with the PDF of the LOS velocity along that direction, which is shown in Fig.~\ref{fig:sim_end_pdfs_001}. We find that the Ideal case exhibits stronger non-Gaussian features than CO. The reason for this is at later times during the collapse, the Ideal case traces all the intermittent features induced by the collapse, while the CO has depleted onto dust grains in the high-density regions and its spatial distribution is such that it probes primarily the envelope of the cloud, where the collapse-induced non-Gaussian features are still largely absent. This is further visualised in the top panels (LOS~1~0~0) of Fig.~\ref{fig:los_late_time}, where we see smaller spatial variations in the LOS velocity in CO compared to the Ideal case. However, at even later times during the collapse we would expect that also the regions further away from the collapse centre, i.e., the envelope would develop more and more of these features, which may then also become more apparent in the CO lines.

\section{The Effect of Optical Depth}
\label{sec:co_depletion}
As discussed in Section~\ref{sec:turb_isolation}, even without depletion, the variation in $\mathrm{H}_2$ density from the outskirts to the centre of the cloud is significantly larger than the corresponding variation in CO abundance. This leads to a stronger central peak in the zeroth-moment maps of the ideal case. This is due to the fact that in the ideal case, the moment maps are weighted by the total gas density, whereas, in the CO case, they are weighted by the CO abundance. At later times, CO depletion also begins to play a role, further enhancing the differences between the ideal and CO cases. While it is true that only a relatively small volume reaches densities of $\sim10^5 \,\mathrm{cm}^{-3}$, these regions can dominate the features in the moment maps of the ideal case. For example, within a magnetic flux tube passing through the centre of the cloud (i.e., at $x = y = 0$), approximately 75\% of the total mass within the flux tube lies within $z = \pm 0.1\,\rm{pc}$ (i.e., the region with density of $\sim10^5\,\rm{cm^{-3}}$). However, ultimately, the differences observed between the CO and ideal cases will be a combination of both abundance variations of CO and/or CO depletion, and optical depth effects.

\begin{figure*}
\def\arraystretch{0.0}
\begin{tabular}{cc}
\includegraphics[width=0.355\linewidth]{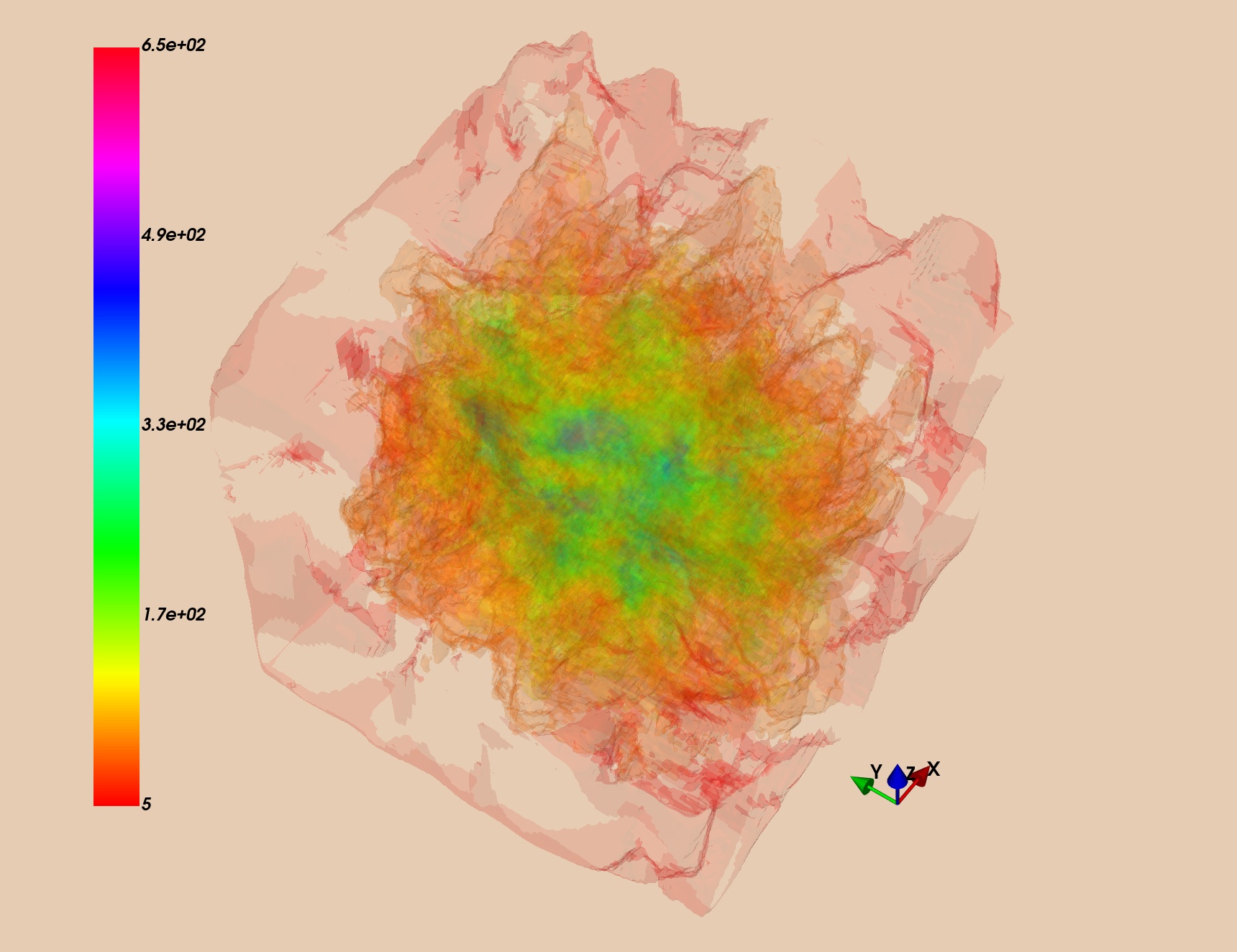} &
\includegraphics[width=0.355\linewidth]{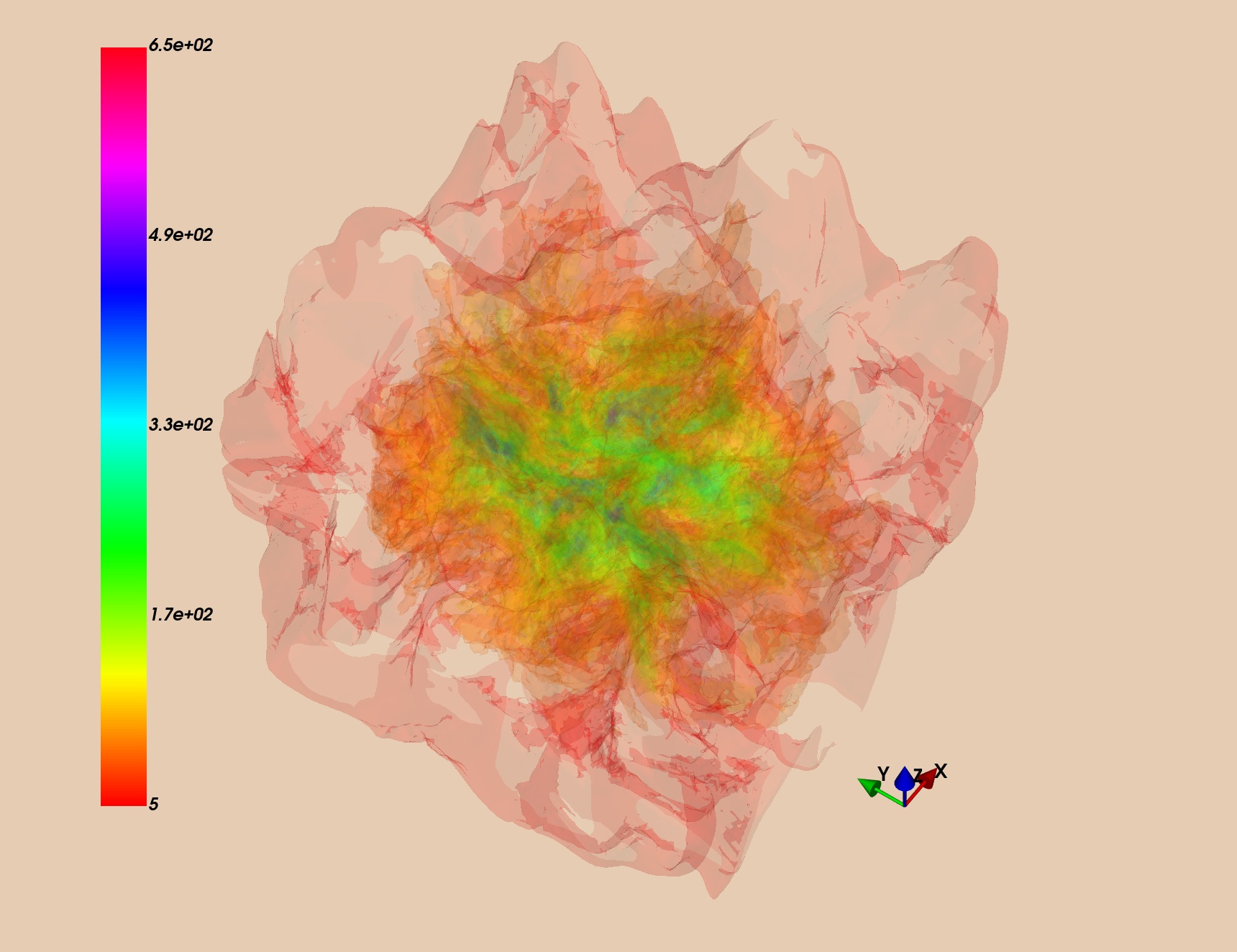}
\end{tabular}
\caption{Iso-surfaces of the optical depth of the $J = 1\rightarrow0$ transition of CO early during the evolution of the cloud (left panel), and at the end of the simulation (right panel). In both instances, the optical depth is very high ($\tau \gg 100$) in the central regions of the cloud.
\label{fig:Odepth}}
\end{figure*}

To decipher which is the driving factor, we present in Fig.~\ref{fig:Odepth} two 3D plots of the optical depth, early (left) and late (right) during the evolution of the cloud, to provide further context. 

In these figures it is evident that the spatial features in the optical depth drastically change between the two instances. In turn, these changes will create differences in the moment maps. However, for both times, the CO line is very optically thick ($\tau \gg 100$) in the densest regions of the cloud.


\bsp	
\label{lastpage}
\end{document}